\documentclass{article}
\usepackage{amsfonts,amssymb, amsmath}
\usepackage[english]{babel}

\def\nn{\nonumber }
\def\bq{ \begin{equation}}
\def\eq{ \end{equation}}
\def\ben{ \begin{eqnarray}}
\def\en{ \end{eqnarray}}

\newtheorem{prop}{Proposition}

\begin{document}

\title{On auto and hetero  B\"{a}cklund  transformations for the H\'{e}non-Heiles
systems}

\author{A. V. Tsiganov\\
\it\small
St.Petersburg State University, St.Petersburg, Russia\\
\it\small e--mail: andrey.tsiganov@gmail.com}

\date{}
\maketitle

\begin{abstract}
We consider a canonical transformation of parabolic coordinates on the plain  and  suppose that this transformation together with some additional relations may be considered as a counterpart of the auto and hetero B\"{a}cklund  transformations associated with  the integrable  H\'{e}non-Heiles systems.

 \end{abstract}

\section{Introduction}
\setcounter{equation}{0}
According to classical definition by Darboux \cite{darb}, a B\"{a}cklund transformation between the two given PDEs
\[E(u,x,t)=0\quad\mbox{and}\quad \tilde{E}(v,y,\tau)=0\]
is a pair of relations
\bq \label{ff-bk}
F_{1,2}(u,x,t,v,y,\tau)=0
\eq
and some additional relations between $(x, t)$ and $(y, \tau)$, which allow to get both equations  $E$ and $\tilde{E}$.
The BT is called an auto-BT or a hetero-BT depending whether the two PDEs are the
same or not.  The hetero-BTs describe a correspondence between equations rather than  a one-to-one mapping between their solutions \cite{ab81,rod02}.  In the modern  theory of partial differential equation B\"{a}cklund  transformations are seen also as a powerful tool in the discretization of PDEs  \cite{bob08}.

 A counterpart of the auto B\"{a}cklund  transformations for finite dimensional integrable systems can be seen as the canonical transformation
\bq\label{bt-coord}
(u,p_u)\to (v,p_v)\,,\qquad \{u_i,p_{u_j}\}=\{v_i,p_{v_j}\}=\delta_{ij}\,,\qquad i,j=1,\ldots,n,
\eq
preserving the algebraic form of the Hamilton-Jacobi equations
\[
  H_i\left(u,\dfrac{\partial S}{\partial u}\right)=\alpha_i\qquad\mbox{and}\qquad {H}_i\left(v,\dfrac{\partial {S}}{\partial v}\right)=\alpha_i
\]
associated with  the Hamiltonians $H_1,\ldots,H_n$ \cite{stef82}.

 The counterpart of the discretization  for finite dimensional  systems is also  currently accepted: by viewing the new $v$-variables as the old $u$-variables, but computed at the next time step; then the B\"{a}cklund  transformation (\ref{bt-coord}) defines  an integrable symplectic map or  discretization of the continuous model, see discussion in \cite{kuz02,zul13}.

   The counterpart of the  hetero  B\"{a}cklund  transformations  for finite dimensional  integrable systems  has to be a canonical transformation (\ref{bt-coord}),  which has to relate  two different systems of the Hamilton-Jacobi equations
   \bq
 \label{Eq-HJt-2} H_i\left(u,\dfrac{\partial S}{\partial u}\right)=\alpha_i\qquad\mbox{and}\qquad \tilde{H}_i\left(v,\dfrac{\partial \tilde{S}}{\partial v}\right)=\tilde{\alpha}_i
\eq
and has to satisfy some additional conditions.  It is necessary to add these conditions to (\ref{bt-coord}) and (\ref{Eq-HJt-2})
in order to get a non-trivial, usable  and efficient theory.  The question of how to do it remains open.

 One of the possible additional conditions may be found in the theory of superintegrable systems. For instance, let us consider integrals of motion for the two-dimensional harmonic oscillator
\[
H_1=p_x^2+p_y^2+a(x^2+y^2)\,,\qquad H_2=p_x^2-p_y^2+a(x^2-y^2)\,,
\]
which yield the Hamilton-Jacobi equations separable in Cartesian coordinates $u=(x,y)$ on the plane. Another pair of Hamiltonians for the same harmonic oscillator
\[
\tilde{H}_1=p_x^2+p_y^2+a(x^2+y^2)\,,\qquad \tilde{H}_2=xp_y-yp_x
\]
is separable in polar coordinates $v=(r,\varphi)$ on the plain. Canonical transformation of variables
\bq\label{ctr-os}
(u,p_u)=(x,y,p_x,p_y)\to(v,p_v)=(r,\varphi,p_r,p_\varphi)
\eq
defines a correspondence between the two different systems of the Hamilton-Jacobi equations
\[
 H_{1,2}\left(x,y,\dfrac{\partial S}{\partial x},\dfrac{\partial S}{\partial y}\right)=\alpha_{1,2}\qquad\mbox{and}\qquad \tilde{H}_{1,2}\left(r,\varphi,\dfrac{\partial \tilde{S}}{\partial r}, \dfrac{\partial \tilde{S}}{\partial \varphi}\right)=\tilde{\alpha}_{1,2}\,.
\]
This  correspondence may be considered as a hetero-BT defined by the generating function
\[
F=p_xr\cos\varphi+p_yr\sin\varphi\,,\]
relations between $(x,y)$ and $(r,\varphi)$
\[
x=r\cos\varphi\,,\qquad y=r\sin\varphi\,,
\]
and  with an additional condition  that Hamilton function $H_1=\tilde{H}_1$  is \textit{simultaneously separable}  in $u$ and $v$ variables.

Canonical transformation (\ref{ctr-os})  can be considered as the semi hetero-BT  relating  different Hamilton-Jacobi equations, which are various faces of the same  superintegrable system.  We know that theory of such  semi hetero-BTs is a profound and very useful theory, both in classical and quantum cases \cite{mil13}.

The main aim of this note is to discuss a correspondence between  integrable H\'{e}non-Heiles systems proposed in \cite{ts14}. This correspondence between different Hamiltonians may be considered as a counterpart of the generic  hetero-BTs relating different but \textit{simultaneously separable}  in  $v$-variables Hamilton-Jacobi equations.

\section{The Jacobi method}
\setcounter{equation}{0}
Let us consider some natural  Hamilton function on $T^*\mathbb R^{n}$
\bq\label{hg-par}
H=p_1^2+\cdots+p_n^2+V(q_1,\ldots,q_n)\,.
\eq
The corresponding Hamilton-Jacobi equation is said to be separable in a set of canonical coordinates $u_i$ if it  has the additively separated complete integral
\[
S(u_1,\ldots,u_n;\alpha_1,\ldots,\alpha_n)=\sum_{i=1}^n S_i(u_i;\alpha_1,\ldots,\alpha_n)\,,
\]
where $S_i$ are found by quadratures as solutions of ordinary differential equations.  In order to express  initial physical variables $(q,p)$ in terms of  canonical  variables of separation $(u,p_u)$ we have to obtain
 momenta $p_{u_i}$ from the second Jacobi equations
\bq\label{2-jeq}
p_{u_i}=\dfrac{ \partial S_i(u_i;\alpha_1,\ldots,\alpha_n)}{\partial u_i}\,,\qquad i=1,\ldots,n.
\eq
Solving these equations with respect to $\alpha_i$ one gets integrals of motion $H_i=\alpha_i$ as functions on  variables of separation $(u,p_u)$.

According to Jacobi we can use canonical transformation $(q,p)\to (u,p_u)$ in order to  construct different integrable systems simultaneously separable in the same coordinates, see pp. 198-199 in  \cite{jac}:\\
"The main difficulty in integrating a given differential equation
lies in introducing convenient variables, which there is no rule for finding.
Therefore, we must travel the reverse path and after finding some notable substitution,
look for problems to which it can be successfully applied."

For instance, let us consider some natural Hamiltonian on the plane
\[
 H_1={p_1^2}+p_2^2 +V(q_1,q_2)
 \]
separable in  the parabolic coordinates $u_{1,2}$
\bq \label{par-coord}
\lambda-2q_2-\dfrac{q_1^2}{\lambda} =\dfrac{(\lambda-u_1)(\lambda-u_2)}{\lambda}\,.
\eq
 In this case second Jacobi equations (\ref{2-jeq}) may be  rewritten in the following form
 \bq\label{par-sep-rel}
p_{u_i}^2+U_i(u_i)=H_1+\dfrac{H_2}{u_i}\,,\qquad i=1,2\,,\qquad
\eq
where  $U_i(u_i)$ are  functions defined by the potential  $V(q_1,q_2)$ \cite{ll}.

Adding together  the separated relations (\ref{par-sep-rel}) one gets another integrable Hamiltonian
\bq\label{ht-gen}
\tilde{H}_1=\dfrac{1}{2}\Bigl( p_{u_1}^2+U_1(u_1)+p_{u_2}^2+U_2(u_2)\Bigr) =H_1+\dfrac{H_2}{2}\left(\dfrac{1}{u_1}+\dfrac{1}{u_2}\right)\,,
\eq
which can be considered as an integrable perturbation of $H$  (\ref{hg-par}) because there exists second independent integrals of motion
\[
\tilde{H}_2=\Bigl( p_{u_1}^2+U_1(u_1)-p_{u_2}^2-U_2(u_2)\Bigr) ={H_2}\left(\dfrac{1}{u_1}-\dfrac{1}{u_2}\right)\,.
\]
 Of course, in generic case this perturbation has no physical meaning.  Auto-BTs of the initial Hamilton-Jacobi equation
\[
(u_i,p_{u_i})\quad \to\quad (v_i,p_{v_i})
\]
preserve  an algebraic form of the initial Hamiltonian $H$  (\ref{hg-par}) and change the form of the second Hamiltonians
 \bq\label{th-par}
\tilde{H}_1=H_1+\dfrac{H_2}{2}\left(\dfrac{1}{v_1}+\dfrac{1}{v_2}\right)\,,\qquad
\tilde{H}_2=H_2\left(\dfrac{1}{v_1}-\dfrac{1}{v_2}\right)\,.
\eq
We can try to pick out a special and maybe unique auto-BT,  which gives  physical meaning to the second Hamiltonian $\tilde{H}$,  as some of the possible counterparts of the hetero-BTs relating different but \textit{simultaneously separable} Hamilton-Jacobi equations.

\subsection{The H\'{e}non-Heiles systems}
There are three  integrable H\'{e}non-Heiles systems on the plane, which can be identified with appropriate finite-dimensional reductions of the integrable  fifth order KdV, Kaup-Kupershmidt  and  Sawada-Kotera equations \cite{fordy91}.
An explicit integration for all these cases is discussed in \cite{cont05}.

We can try to get a hetero-BT for the finite-dimensional H\'{e}non-Heiles systems  taking  the hetero-BT between these integrable PDEs  and then applying  the Fordy finite-dimensional reduction \cite{fordy91}. We believe that the same information may be directly extracted from the well-known  Lax representation for the H\'{e}non-Heiles system separable in parabolic coordinates.

Let us take a Lax  matrix for the first H\'{e}non-Heiles system separable in parabolic coordinates
\bq\label{par-lax1}
 L(\lambda)=\left(
 \begin{array}{cc}
 \dfrac{p_2}{2}+\dfrac{p_1q_1}{2\lambda} &\lambda-2q_2-\dfrac{q_1^2}{\lambda} \\ \\
 a\lambda^2+2aq_2\lambda+a(q_1^2+4q_2^2)+\dfrac{p_1^2}{4\lambda} &- \dfrac{p_2}{2}-\dfrac{p_1q_1}{2\lambda} \\
 \end{array}
 \right)\,,\qquad  a\in\mathbb R\,\,.
 \eq
Characteristic polynomial of this  matrix
\[
\det\Bigl(L(\lambda)-\mu\Bigr)=\mu^2-a\lambda^3-\dfrac{H_1}{4}+\dfrac{H_2}{\lambda}
\]
contains the Hamilton function of the first H\'{e}non-Heiles system associated with a  fifth order KdV
\bq\label{hh-1}
H_1=p_1^2+p_2^2-16aq_2(q_1^2+2q_2^2)
\eq
and a second integral of motion
\bq\label{hh-12}
H_2=aq_1^2(q_1^2+4q_2^2)+\dfrac{p_1(q_2p_1-q_1p_2)}{2}\,.
\eq

The auto-BTs  preserve the algebraic form of the Hamiltonians \cite{stef82}. Since the characteristic polynomial is the generating function of these integrals of motion, their invariance amounts to requiring the existence of a  similarity transformation for  the Lax matrix
\[\hat{L}=V{L}V^{-1}\,,\]
associated with the given auto-BT. The matrix $V$ needs not to be unique because a dynamical system can
have different auto BTs  \cite{kuz02,zul13}.

In contrast with  \cite{kuz98} we do not require that the transformed Lax matrix $\hat{L}(\lambda)$ have the same structure
in the spectral parameter $\lambda$ as the original Lax matrix. This requirement is a property of the particular BTs  associated with the special translations on hyperelliptic Jacobians.

Let us consider a special, unique similarity transformation associated  with matrix
 \bq\label{v-mat} V=\left(
 \begin{array}{cc}
 {L}_{12} & 0 \\
 4\bigl({L}_{11}-\hat{L}_{11}(\lambda)\bigr )& 4{L}_{12}\\
 \end{array}
 \right)\,,\eq
where $L_{ij}$ are entries of the Lax matrix (\ref{par-lax1}) and
\[
\hat{L}_{11}(\lambda)=\dfrac{p_2}{2}+\dfrac{p_1(\lambda-2q_2)}{2q_1}\,.
\]
The    Lax matrix $\hat{L}(\lambda)=V{L}V^{-1}$ has the following properties:
\begin{enumerate}
  \item first off-diagonal element of the Lax matrix
  \[\hat{L}_{12}(\lambda)=\dfrac{(\lambda-u_1)(\lambda-u_2)}{4\lambda}\]
  yields initial parabolic coordinates on the plane (\ref{par-coord});
  \item  second off-diagonal element
 \ben\label{v-hh}
\hat{L}_{21}&=&4a{(\lambda-v_1)(\lambda-v_2)}\\ \nn\\
&=&4a\lambda^2+\dfrac{(8aq_1^2q_2-p_1^2)\lambda}{q_1^2}+
4a(q_1^2+4q_2^2)+\dfrac{2p_1(p_1q_2-p_2q_1)}{q_1^2} \nn
\en
has only  two commuting and functionally independent zeroes $v_{1,2}$;
\item
the conjugated momenta for $u$ and $v$ variables are  the values of the diagonal element
\[
p_{u_i}=\hat{L}_{11}(\lambda=u_i)\,,\qquad
p_{v_i}=\hat{L}_{11}(\lambda=v_i)\,,\qquad i=1,2.
\]
\end{enumerate}
In generic case such $2\times2$ Lax matrices $\hat{L}(\lambda)$ exist only  if the genus of hyperelliptic curve defined by equation \[\det\Bigl(L(\lambda)-\mu\Bigr)=0\]
 is no more a number of degrees of freedom. The corresponding transformation of the classical $r$-matrix is discussed in \cite{ts14}.

 In $(u,p_u)$ and $(v,p_v)$ variables entries of the  transformed Lax matrix $\hat{L}$ have  the following  form:
\ben
\hat{L}_{11}&=&\dfrac{\lambda-u_2}{u_1-u_2}p_{u_1}+\dfrac{\lambda-u_1}{u_2-u_1}p_{u_2}
=\dfrac{\lambda-v_2}{v_1-v_2}p_{v_1}+\dfrac{\lambda-v_1}{v_2-v_1}p_{v_2}\nn\\
\nn\\
\hat{L}_{12}&=&\dfrac{(\lambda-u_1)(\lambda-u_2)}{4\lambda}\nn\\
&=&\dfrac{\lambda^2+\lambda(v_1+v_2)+v_1^2+v_1v_2+v_2^2}{4\lambda}-\dfrac{(p_{v_1}-p_{v_2})^2}{4a(v_1-v_2)^2}
-\dfrac{p_{v_1}^2-p_{v_2}^2}{4\lambda(v_1-v_2)}
\nn\\ \nn\\
\hat{L}_{21}&=&
4a\bigl(\lambda^2+\lambda(u_1+u_2)+u_1^2+u_1u_2+u_2^2\bigr)
-\dfrac{4\lambda(p_{u_1}-p_{u_2})^2}{(u_1-u_2)^2}-\dfrac{4(p_{u_1}^2-p_{u_2}^2)}{u_1-u_2}\nn\\
&=&4a{(\lambda-v_1)(\lambda-v_2)}\nn
\en
Thus, we have two families of variables of separation for the  first H\'{e}non-Heiles system and canonical transformation between them:
\ben
u_{1,2}&=&-\dfrac{v_1+v_2}{2}+\dfrac{ (p_{v_1}-p_{v_2})^2\pm\sqrt{A} }{2a(v_1-v_2)^2}\,,
\nn\\
\label{bt-hh}\\
p_{u_{1,2}}&=&\dfrac{(p_{v_1}-p_{v_2})\bigl( (p_{v_1}-p_{v_2})^2 \pm \sqrt{A}\,\bigr)}{2a(v_1-v_2)^3}
-\dfrac{p_{v_1}(v_1+3v_2)-p_{v_2}(v_2+3v_1)}{2(v_1-v_2)}\nn
\en
where
\ben
A&=&(p_{v_1}-p_{v_2})^4+2a(v_1-v_2)^2(p_{v_1}-p_{v_2})\bigl(p_{v_1}(v_1-3v_2)-p_{v_2}(v_2-3v_1)\bigr)\nn\\
\nn\\
&-&a^2(3v_1^2+2v_1v_2+3v_2^2)(v_1-v_2)^4\,.\nn
\en
 We can directly prove that Hamiltonains $H_{1,2}$ (\ref{hh-1}-\ref{hh-12})  have the same algebraic form in $(u,p_u)$ and $(v,p_v)$ variables using this explicit canonical transformation.

\begin{prop}
The auto-BT for the first H\'{e}non-Heiles system is  a correspondence between two equivalent systems of the Hamilton-Jacobi equations
\[
H_{1,2}\left(\lambda,\dfrac{\partial S}{\partial \lambda}\right)=\alpha_{1,2}\,,\qquad \lambda=u,v\,,
\]
where variables $(u,p_u)$ and $(v,p_v)$ are related by canonical transformation  (\ref{bt-hh}) and
 Hamiltonians  $H_{1,2}$ are defined by the following equations
\bq\label{c-par1}
\Phi(\lambda,\mu)=\mu^2-a\lambda^3=\dfrac{H_1}{4}-\dfrac{H_2}{\lambda}\,,\qquad \lambda=u_{1,2},v_{1,2},\quad \mu=p_{u_{1,2}},p_{v_{1,2}}\,.
\eq

 \end{prop}
This auto B\"{a}cklund  transformation changes coordinates on an algebraic invariant manifold defined by $H_{1,2}$ without changing the manifold itself \cite{kuz02}.

We can convert this special, unique auto-BT to some analogue of the hetero-BT by adding  one more relation.
Namely,  substituting roots of the off-diagonal element $\hat{L}_{2,1}$ (\ref{v-hh}) into the definition (\ref{th-par})
one gets the Hamilton function for the second integrable H\'{e}non-Heiles system associated with the Kaup-Kupershmidt equation
\bq\label{hh-2}
\tilde{H}_1=p_1^2+p_2^2-2aq_2(3q_1^2+16q_2^2)
\eq
up to rescaling  $p_1\to \sqrt{2}p_1$ and $ q_1\to q_1/\sqrt{2}$.

After canonical transformation
\bq\label{v-PQ}
(q,p)\to (Q,P)\,,\qquad P_{1,2}=\dfrac{p_{v_1}\pm p_{v_2}}{\sqrt{2}}\,,\quad Q_{1,2}=\dfrac{v_1\pm v_2}{\sqrt{2}}
\eq
the same  Hamiltonian
\bq\label{hh-3}
\tilde{H}_1=P_1^2+P_2^2-2aQ_2(3Q_1^2+Q_2^2)
\eq
defines a third  integrable H\'{e}non-Heiles system associated with the  the Sawada-Kotera equation. According \cite{fordy91} canonical transformation (\ref{v-PQ}) is a counterpart of the gauge equivalence of the Sawada-Kotera and Kaup-Kupershmidt equations.

So, all the  H\'{e}non-Heiles systems on the plane are \textit{ simultaneously separable} in $v$-variables, and we suppose that this fact allows us to  define  some natural counterpart of the hetero-BT.
\begin{prop}
For the H\'{e}non-Heiles systems on the plane  (\ref{hh-1}) and  (\ref{hh-2},\ref{hh-3}) an analogue of the hetero-BT is the correspondence between two different systems of the Hamilton-Jacobi equations
\[
  H_{1,2}\left(u,\dfrac{\partial S}{\partial u}\right)=\alpha_{1,2}\qquad\mbox{and}\qquad \tilde{H}_{1,2}\left(v,\dfrac{\partial \tilde{S}}{\partial v}\right)=\tilde{\alpha}_{1,2}\,,
\]
where variables $(u,p_u)$ and $(v,p_v)$ are related by canonical transformation  (\ref{bt-hh}) and
 Hamiltonians are defined by the following equations
\[
\Phi(\lambda,\mu)=\mu^2-a\lambda^3=\dfrac{H_1}{4}-\dfrac{H_2}{\lambda}\,,\qquad \lambda=u_{1,2},v_{1,2},\quad \mu=p_{u_{1,2}},p_{v_{1,2}}\,,
\]
 and
 \[\tilde{H}_{1,2}=\Phi(v_1,p_{v_1})\pm \Phi(v_2,p_{v_2})\,.
  \]

 \end{prop}
This analogue of the hetero-BT relates different algebraic invariant manifolds associated with Hamiltonians $H_{1,2}$ and $\tilde{H}_{1,2}$ similar to the well-studied relations between different invariant manifolds in the theory of superintegrable systems.

For the first H\'{e}non-Heiles system on the plane  (\ref{hh-1})  we can consider parabolic variables $(u_{1,2},p_{u_{1,2}})$ as coordinates  on the Jacobian variety defined by equations (\ref{c-par1}). In order to get integrals of motion for the second or third H\'{e}non-Heiles systems (\ref{hh-2},\ref{hh-3}) we have to take linear combinations of these equations and to make simultaneously the special shift of the coordinates $(u,p_u)\to(v,p_v)$ on the Jacobian variety.

 \section{Integrable Hamiltonian with velocity dependent potential}
 \setcounter{equation}{0}
 It is well-known that Hamilton-Jacobi equation is separable in parabolic coordinates $u_{1,2}$ if the Hamilton function has the form
\[
H=p_1^2+p_2^2+V_N(q_1,q_2)\,,\qquad
 V_N=4a\sum_{k=0}^{[N/2]} 2^{1-2k}\left(
                                  \begin{array}{c}
                                    N-k \\
                                    k \\
                                  \end{array}
                                \right) q_1^{2k}q_2^{N-2k}\,,
\]
where the positive integer $N$ enumerates the members of the hierarchy.

At $N=3$ one gets the  H\'{e}non-Heiles system (\ref{hh-1}), at $N=4$ the next member of hierarchy
is a "(1:12:16)" system with the following Hamiltonian
\bq\label{dd-ham}
H=p_1^2+p_2^2-4a\left(q_1^4+12q_1^2q_2^2+16q_2^4\right)\,.
\eq
The corresponding Lax matrix is equal to
\bq\label{par-lax2}
 L(\lambda)=\left(
 \begin{array}{cc}
 \dfrac{p_2}{2}+\dfrac{p_1q_1}{2\lambda} &\lambda-2q_2-\dfrac{q_1^2}{\lambda} \\ \\
 a\lambda^3+2aq_2\lambda^2+a(q_1^2+4q_2^2)\lambda+4aq_2(q_1^2+2q_2^2)+\dfrac{p_1^2}{4\lambda} &- \dfrac{p_2}{2}-\dfrac{p_1q_1}{2\lambda} \\
 \end{array}
 \right)\,.
 \eq
After  similarity transformation of $L(\lambda)$ with matrix $V$ (\ref{v-mat}), where
\[
\hat{L}_{11}(\lambda)=\sqrt{a}\,\lambda^2-\dfrac{4\sqrt{a}q_2q_1-p_1}{2q_1}\,\lambda-\dfrac{2\sqrt{a}q_1^3+2p_1q_2-p_2q_1}{2q_1}\,,
\]
one gets the transformed Lax matrix with two off-diagonal elements $\hat{L}_{12}(\lambda)$ and $\hat{L}_{21}(\lambda)$, which yield two families of variables of separation.

As above first coordinates are  parabolic coordinates $u_{1,2}$, whereas second coordinates $v_{1,2}$ are zeroes of the polynomial
 \ben\label{v-22}
\hat{L}_{21}&=&\dfrac{4(4aq_1q_2-\sqrt{a}p_1)}{q_1}\,\lambda^2+\dfrac{8aq_1^2(q_1^2+2q_2^2)
+4\sqrt{a}q_1(2p_1q_2- p_2 q_1)-p_1^2}{q_1^2}\,\lambda\nn\\
\nn\\
&+&\dfrac{16 a q_1^2 q_2(q_1^2+2 q_2^2)+2p_1(p_1 q_2- p_2 q_1)}{q_1^2}=\dfrac{4(4aq_1q_2-\sqrt{a}p_1)}{q_1}\,(\lambda-v_1)(\lambda-v_2)\,.\nn
\en
Substituting roots of this polynomial into the definition
(\ref{th-par}) one gets integrable Hamiltonian with velocity dependent potential
\[
\tilde{H}=\dfrac{p_1^2}{2}+p_2^2+4\sqrt{a}p_1q_1q_2-2\sqrt{a}p_2q_1^2-8aq_2^2(5q_1^2+8q_2^2)\,.
\]
Using  canonical transformation  we can rewrite this Hamiltonian in a more symmetric form
\bq\label{tH-vel}
\tilde{H}=p_1^2+p_2^2-3\sqrt{a}p_2q_1^2+2a(q_1^4-12q_1^2q_2^2-32q_2^4)\,.
\eq
The corresponding second integral of motion is  fourth order polynomial in momenta
\ben
\tilde{H}_2&=&p_1^4+4q_1^4\left(q_1^4-8q_1^2q_2^2-112q_2^4\right)a^2+4q_1^3\left(64p_1q_2^3-p_2q_1^3-12p_2q_1q_2^2\right)a^{3/2}\nn\\
\nn\\
&+&q_1^2\left(4p_1^2q_1^2-48p_1^2q_2^2+32p_1p_2q_1q_2+p_2^2q_1^2\right)a-6a^{1/2}p_1^2p_2q_1^2\,,\nn
\en
which also can be obtained from the Lax matrix $\hat{L}(\lambda)$. Of course, this integrable system on the plane (\ref{tH-vel}) could be obtained in the framework of  different theories, see \cite{fordy91,puc04,yeh13} and references within.

Canonical transformation (\ref{v-PQ}) allows us to identify a Hamilton function  with  velocity dependent potential (\ref{tH-vel}) and Hamilton function
\[
\hat{H}=P_1^2+P_2^2-a(Q_1^4+6Q_1^2Q_2^2+Q_2^4)
\]
similar to  the relation between second and third  H\'{e}non-Heiles systems.

Canonical transformation $(u,p_u)\to (v,p_v)$ is the special auto-BT  for the "(1:12:16)"\, system, which can be considered as a hetero-BT relating two different Hamilton-Jacobi equations associated with Hamiltonians $H$ (\ref{dd-ham}) and $\tilde{H}$ (\ref{tH-vel}), respectively.

\section{Conlusion}
The problem of finding separation coordinates for the Hamilton-Jacobi equations is highly non-trivial. The problem was originally formulated by Jacobi when he invented elliptic coordinates and successfully applied them to solve several
important mechanical problems with quadratic integrals of motion in momenta.

We suppose that  after  suitable B\"{a}cklund transformations standard elliptic, parabolic  etc. coordinates   turn into
 variables of separation  for physically interesting  integrable systems with higher order integrals of motion.  For example,
 in this note we have constructed a canonical transformation of the standard parabolic coordinates, which yields variables of separation for the three integrable H\'{e}non-Heiles systems.

 Moreover, we believe that information about such suitable B\"{a}cklund transformations and the corresponding integrable systems is  incorporated into the  Lax matrices associated with these elliptic, parabolic  etc. coordinates. In order to prove it we obtained  integrals of motion,  variables of separation and separated relations for some new integrable system with velocity dependent potential and fourth order integral of motion in momenta.  In  similar manner we can construct various simultaneously separable integrable systems associated with other curvilinear coordinates on the Riemannian manifolds of constant curvature, see examples in \cite{ts14,ts15}.

\vskip0.2truecm
\par\noindent
This work was partially supported by RFBR grant 13-01-00061 and SPbU grant 11.38.664.2013.

\end{document}